\documentclass[pra, final, showpacs, twocolumn, groupedaddress, amsfonts, floatfix]{revtex4}

\usepackage{amsmath, isolatin1, graphicx, color, amsxtra, amssymb, bm,
  dcolumn}

\newcommand{\ket}[1]{| #1 \rangle}
\newcommand{\bra}[1]{\langle #1 |}

\newcommand{\ud}{\text{d}}
\newcommand{\refeqn}[1]{Eq.~(\ref{#1})}
\newcommand{\reffig}[1]{Fig.~\ref{#1}}
\newcommand{\reftab}[1]{Table~\ref{#1}}
\newcommand{\ph}{{\phantom{1}}}

\newcommand{\ts}{\text{s}}
\newcommand{\ti}{\text{i}}
\newcommand{\tp}{\text{p}}
\newcommand{\tV}{\text{V}}
\newcommand{\tH}{\text{H}}
\newcommand{\tD}{\text{D}}
\newcommand{\tA}{\text{A}}
\newcommand{\tg}{\text{g}}
\newcommand{\tc}{\text{c}}

\newcommand{\etal}{\emph{et~al.\ }}
\newcommand{\prodrateunit}[1]{{#1} s$^{-1}$THz$^{-1}$mW$^{-1}$}

\begin{document}

\title{Theory and experiment of entanglement in a quasi-phase-matched
  two-crystal source}

\author{Daniel Ljunggren}
\email[Corresponding author. Electronic address: ]{daniellj@kth.se}
\homepage{http://www.quantum.se}
\author{Maria Tengner}
\author{Philip Marsden}
\altaffiliation[Current address: ]{Department of Physics, University
  of Toronto, Toronto M5S 1A7, Canada}
\author{Matthew Pelton}
\altaffiliation[Current address: ]{Department of Physics, University
  of Chicago, Chicago, IL 60637, USA}
\affiliation{Department of Microelectronics and
  Information Technology, The Royal Institute of Technology, KTH,
  Electrum 229, SE-164 40 Kista, Sweden}

\date{October 8, 2005}

\begin{abstract}
  We report new results regarding a source of polarization entangled
  photon-pairs created by the pro\-cess of spontaneous parametric
  downconversion in two orthogonally oriented, periodically poled,
  bulk {KTiOPO$_4$} crystals (PPKTP).  The source emits light
  colinearly at the non-degenerate wavelengths of 810 nm and 1550 nm,
  and is optimized for single-mode optical fiber collection and
  long-distance quantum communication. The configuration favors long
  crystals, which promote a high \mbox{photon-pair} production rate at
  a narrow bandwidth, together with a high pair-probability in fibers.
  The quality of entanglement is limited by chromatic dispersion,
  which we analyze by determining the output state.  We find that such
  a decoherence effect is strongly material dependent, providing for
  long crystals an upper bound on the visibility of the coincidence
  fringes of 41\% for {KTiOPO$_4$}, and zero for {LiNbO$_3$}.  The
  best obtained raw visibility, when canceling decoherence with an
  extra piece of crystal, was $91 \pm 0.2\%$, including background
  counts. We confirm by a violation of the \mbox{CHSH-inequality} (${S
    = 2.679 \pm 0.004}$ at 55 s$^{-1/2}$ standard deviations) and by
  complete quantum state tomography that the fibers carry high-quality
  entangled pairs at a maximum rate of $55 \times 10^3$
  s$^{-1}$THz$^{-1}$mW$^{-1}$.

\end{abstract}


\maketitle

\section{Introduction}

A nonlinear medium exposed to an optical field will occasionally emit
several other photons. The pheno\-menon is known as spontaneous
parametric downconversion (SPDC), and is frequently utilized for the
production of photon-pairs.  Such a pair can also become entangled in
a certain degree of freedom if indistinguishability is ensured in all
the remaining degrees of freedom.  Many successful examples of direct
creation of entangled photon-pairs \cite{KMWZSS95, JSWWZ00, KOW01a},
post-selected entangled pairs \cite{KSSA93, TBGZ99, KKS00}, and
in-fiber generated pairs \cite{TI04, LVCSK05, LVSK05} can be given,
already serving as an indispensable tool for quantum communication.

The source reported here uses two orthogonally oriented crystals,
each emitting pairs of photons of a different polarization than the
other. The different pairs are made indistinguishable, in our case
by single-mode fibers, and therefore the individual photons of a
single pair become directly entangled in polarization --- an idea
originally proposed by Hardy \cite{Hardy92} and realized in modified
form by Kwiat \etal \cite{KWWAE99}. One problem with the original
realization is that the crystals cannot be made too long, since the
non-colinearity makes the two emission-cones non-overlapping.
Another problem is that the crystals used generally emit into many
spatial modes, which is not suitable for fiber-coupling.  Using
periodically poled crystals via quasi-phase matching \cite{FMKWS04,
MAKW02, TTRZBMOG02}, it has been shown that colinear emission can be
achieved very close to a single mode \cite{LT05} (even in
non-waveguiding structures), providing much greater overlap in the
emission.  Such a configuration also allows non-degenerate
wavelengths to be generated.

Some desirable properties of sources to be used for quantum
communication include: i) a high probability of photon-pairs to be
collected into optical fibers; ii) a minimum number of false
coincidences; iii) wavelength combinations that either suit
efficient detection, match atomic transitions, or are well
transmitted over long distances; iv) a narrow bandwidth that limits
the effects of fiber dispersion \mbox{($\sim$\ GHz)} \cite{HTRBZG05}
or can address atoms \mbox{($\sim$\ MHz)}; v) a long coherence
length that limits the need for precise interferometry; vi) small
jitter in arrival-time of photons; vii) perfect correlations in all
bases; and, ideally, viii) the source being compact enough to be put
in a box, carried out of the lab, and be used, e.g.,\ for quantum
key distribution (QKD). Furthermore, for maximum security in QKD a
strong requirement is to have neither more nor less than a single
photon per gate pulse. In this respect, photon-pair sources have
been shown to be good candidates compared to weak coherent pulses,
potentially fulfilling properties i) and ii). Equally imperative for
security in Ekert's scheme \cite{Ekert91} is property vii), which
expresses the wish for high visibility of entanglement in the
presence of background detection, which implies the need to minimize
dark counts and false coincidence counts.

In this work, we extend our previous results \cite{PMLTKFCL04}
regarding a PPKTP-based two-crystal source and try to address some
of the anticipated features above.  By emitting at non-degenerate
wavelengths, the source exploits the highly efficient Si-based
single-photon counters available in the near-infrared region and the
low attenuation in fibers at telecom wavelengths. The shorter
wavelength also matches the transmission bands of alkaline atoms,
which makes the source suitable as part of a quantum memory
\cite{TTHABGZ05,
  KMWA05}.  For our crystal configuration, we show how effects like
chromatic dispersion enter the picture as problems to be dealt with.
The source has been optimized for coupling into single-mode fibers
following Ref.\ \cite{LT05}, where one can also find motivations for
using long crystals to achieve a narrow bandwidth.  An early example
of a non-degenerate source is Ref.\ \cite{RBGGZ01}, using
energy-time entanglement. One reason for utilizing energy-time
entanglement is to overcome the strong decoherence mechanism of
polarization-states over fibers, and, for the same reason, we
propose a scheme that combines time-multiplexed encoding on the
telecom wavelength side \cite{TBZG00} with polarization on the
near-infrared side, altogether realizing a sort of hybrid-coded
entanglement.

The article is organized as follows. In section II, we describe the
main characteristics of the source. In section III, we derive the
quantum state emitted by the two crystals in terms of frequency and
polarization degrees of freedom, based on the quantum state of a
single crystal derived in the Appendix. Following that, in section
IV, we briefly show how to compensate for the effect of chromatic
dispersion in the crystals, so as to assure indistinguishability,
and, in section V, we present our experimental results showing the
quality of the source, including results on quantum state
tomography.  In section VI, we discuss the future directions of a
hybrid-coded source, and we end with a summary in section VII.

\section{A source of polarization entanglement}

\begin{figure}[t!]
    \begin{center}
      \includegraphics[scale = 1]{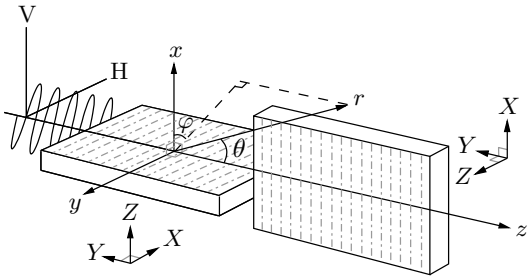}
    \caption{The source consists of two PPKTP crystals placed
      one after the other; the first creates a vertically
      (V) and the second a horizontally (H) polarized field. The laboratory coordinate
      system is drawn, as well as the crystal axes, $X$, $Y$, and
      $Z$, which refer to the polarization of the incoming and
      outgoing electromagnetic fields.}
    \label{Fig:sourcecoordinate}
    \end{center}
\end{figure}
The source is depicted in Fig.~\ref{Fig:sourcecoordinate}, and
consists of two orthogonally aligned bulk crystals placed one after
the other. They each have the dimensions ${3\times4.5\times1}$ mm
(${X,Y,Z}$), of which the second dimension defines the length, $L =
4.5$ mm. The crystals are made of potassium titanyl phosphate,
{KTiOPO$_4$}, and are periodical poled with the period $\Lambda =
9.6\ \mu$m, chosen such that we have phase-matching for the signal
at a wavelength of 810 nm, and the idler at 1550 nm, for a
temperature $T = 111^\circ$C determined by the Sellmeier equations
of \mbox{KTP} \cite{FHHEFBF87, FASR99}.  The crystals are pumped by
monochromatic and continuous wave laser light (p) at a wavelength of
532 nm, which is propagating in a Gaussian TEM$_{00}$ mode along the
$z$-axis, producing a signal (s) and idler (i) field in the same
direction and with the same polarization as the $Z$-component of the
pump field ($Z_\tp Z_\ts Z_\ti$).  \reffig{Fig:sourcecoordinate}
defines the laboratory axes and the crystals' optical axes $X$, $Y$,
and $Z$, oriented as shown. Both crystals will generate
downconverted light if the pump polarization is oriented at
$45^\circ$ to the horizontal (H) and vertical (V) axes.  Following
\cite{LT05}, we have optimized the focusing of the pump and the
fiber-matched modes using the parameter $\xi = L / z_R$, where $L$
is the length of the crystal and $z_R$ is the Rayleigh range, such
that the maximum amount of emission that is generated is collectible
into single-mode fibers. The optimal values for our configuration
are $\xi_\tp = 1.3$, $\xi_\ts = 2.0$, and $\xi_\ti = 2.3$,
respectively.

The use of single-mode fibers to collect the light will erase all
spatial information that reveals from which crystal the photons
came, except for the polarization degree of freedom. Therefore, each
of the beams will interfere in the diagonal basis and get entangled
in polarization. (Note that the spatial information is partly
correlated with frequency via the phase-matching condition, and that
indistinguishability could also be achieved via frequency
filtering.) The resulting state is the Bell-state,
\begin{align}
  \label{Eqn:bell-state}
  \ket{\Phi^\varphi} =
  \frac{1}{\sqrt{2}}\left(\ket{\tV}_\ts\ket{\tV}_\ti +
    e^{i\varphi}\ket{\tH}_\ts\ket{\tH}_\ti\right),
\end{align}
with a relative phase $\varphi$ that we can control. As in most
cases, it is required that the probability of creating more than a
single pair within a time determined by the coherence time of the
photons, or the detector gate-time, whichever is longer, is
negligible, and for moderate pump-powers and relatively short
gate-times or wide bandwidths, this probability is very small, but
not vanishing.  Assuming a Poissonian distribution, the probability
becomes $P_{n \geq 2} = 1 - (1 + m) e^{-m}$, where $m = \Delta
t_\text{g} \beta P_\tp \lambda_\tp /hc$ is the mean photon number in
a single random gating. For a typical detector gate-time $\Delta
t_\text{g} = 5$ ns, pump-power $P_\tp = 540\ \mu\text{W}$, and
conversion efficiency $\beta = 3 \times 10^{-10}$ we get $m = 2
\times 10^{-3}$ and $P_{n \geq 2} = 2 \times 10^{-6}$.

\begin{figure}[tb!]
  \begin{center}
    \includegraphics[scale = 1]{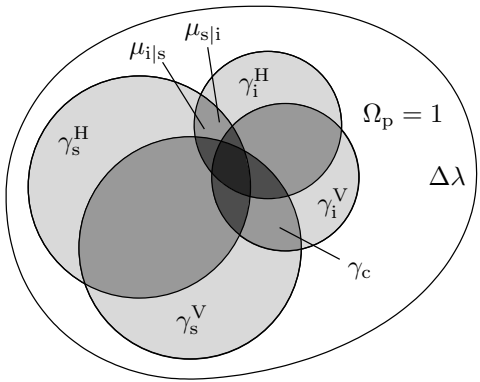}
    \caption{The figure shows a Venn diagram, which is used to
      illustrate the single coupling efficiencies $\gamma_\ts$ and
      $\gamma_\ti$, pair coupling $\gamma_\text{c}$, and conditional
      coincidences $\mu_{\ts |\ti}$ and $\mu_{\ti | \ts}$ as a
      fractional number representing the area of a set. The elements
      contained in a specific set represent photon pairs that are
      coupled into a fiber taken from the universal set of pairs,
      $\Omega_\tp$, which contains all pairs generated from the
      crystals (V or H) within the bandwidth of the detector filter
      $\Delta\lambda$.  The set $\Omega_\tp$ is normalized to unity
      and represents perfect coupling. Maximum overlap of all sets is
      needed to generate the best entanglement in the fiber, which is
      represented by the darkest shaded area in the diagram (the
      union of all sets).}
    \label{Fig:venn_diagram}
  \end{center}
\end{figure}

\reffig{Fig:venn_diagram} will serve as an illustration of the
problem of optimizing the focus of the pump-mode, and the
fiber-matched modes with respect to \emph{two} crystals. As a
compromise, the pump-beam is focused at the interface between the
crystals, in the anticipation that the profile of the generated
emission exactly trails the profile of the pump-beam.  However,
numerical simulations with the software developed in \cite{LT05}
show that the waist of the emission will be shifted towards the
center of each crystal, so that neither the vertically nor the
horizontally polarized photons will couple perfectly into the fiber
simultaneously. The figure shows the different types of coupling
efficiencies represented as sets in a Venn-diagram, where each
element of a set represents a photon pair generated by the crystals
in some spatial mode. That is, the collection of all elements within
each set defines which pairs are coupled into the fiber for some
specific focusing condition, in such a way that the coupling
efficiency corresponds to the total area of the set. The problem can
be described in two parts: first, the need to overlap the matching
modes of the signal and idler, represented by the coupling
efficiencies $\gamma_\ts$ and $\gamma_\ti$, for each polarization
separately (i.e.\ by optimizing the pair coupling $\gamma_\tc =
\mu_{\ti |\ts} \gamma_\ts$, via the conditional coincidence
$\mu_{\ti |\ts}$), and second, the need to overlap the vertically,
$\gamma^\tV$, and the horizontally, $\gamma^\tH$, polarized photons
for both the signal and idler. It is only in the intersection of all
sets where entanglement exits, and any detection of photons outside
of this set will limit the visibility in the $\pm 45^\circ$-basis
(denoted here D/A-basis) by contributing to a mixed state. This
picture is valid for many types of sources, and we believe that the
coupling efficiencies in many cases in the literature are estimated
in an incorrect way, as it is important to note that $\gamma_\tc
\neq \gamma_\ts \gamma_\ti$ (especially in non-degenerate regimes).
By this short discussion (see \cite{TL05} for a comprehensive
discussion), we hope to have illustrated that it is not necessarily
best to optimize each arm individually to find the greatest
coincidences, but rather, to simultaneously optimize both arms.

\begin{figure}[tb!]
    \begin{center}
    \includegraphics[scale = 0.8]{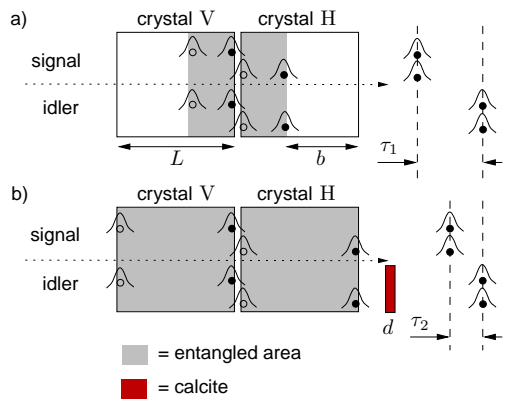}
    \caption{Color online. The figure illustrates the effects of
      chromatic dispersion resulting from the strong non-degeneracy of
      the signal and idler photons, with group refractive indices
      $n_{\tg,\ts}$ and $n_{\tg,\ti}$. a) A pair created at the
      end-facet of the V-crystal (black dots) will non-interactively
      pass through the H-crystal, after which the signal and idler
      wavepackets become separated by a time $\tau_1 = (n_{\tg,\ts}^X
      - n_{\tg,\ti}^X)L/c$ before hitting the detectors. As seen from
      the detectors' viewpoint, for the same pair to instead have been
      created in the H-crystal (black dots), its wavepackets would
      necessarily need to have separated by the same amount, given by
      $\tau_1^\prime = (n_{\tg,\ts}^Z - n_{\tg,\ti}^Z)b/c$, in order
      to interfere with (i.e.\ overlap with, or be indistinguishable
      from) the first case.  (The superscripts refers to different
      polarization-axes.)  Any pair created within the gray area
      (between black and white dots), are separated by a time $\tau
      \geq \tau_1 = \tau_1^\prime$, and will always find a
      corresponding position in the orthogonal crystal to interfere
      with, according to the detectors; however, all the pairs from
      the white area of either crystal will be \emph{distinguishable}
      in time from any pair of the other crystal, i.e.\ the
      wavepackets are non-overlapping due to different dispersions,
      and contribute therefore to a mixed output state.  b) If we put
      a birefringent plate of thickness $d$ in one of the arms, the
      time-separation for a pair created at the end-facet of the
      V-crystal is reduced to $\tau_2 = (n_{\tg,\ts}^X -
      n_{\tg,\ti}^X)L/c + (1 - n_{\tg,\tc}^e)d/c < \tau_1$, for which
      some $d$ equals the time-separation of an interfering position
      at the \emph{end-facet} of the H-crystal, $\tau_2 = -(1 -
      n_{\tg,\tc}^o)d/c$.  Consequently, all H and V-pairs now show
      ``self-interference'', and a pure output state is created.  Note
      that two pairs created within the coherence time of the pump
      (which needs to be longer than $L$) are always coherent,
      ignoring dispersion.}
    \label{Fig:intuitive}
    \end{center}
\end{figure}
As we have mentioned, the different polarizations need to interfere,
and therefore a major concern is that they are not distinguishable
by time information, noting the limited extent of the photon
wave-packets. For long crystals, the photon pairs will separate by
chromatic dispersion, due to the very different group velocities
between the strongly non-degenerate signal and idler. This will
occur to a degree that is different for pairs created in the first
crystal than for pairs created in the second, because the pairs from
the first crystal also need to pass through the second.  The
differences in group velocities between signal and idler are not the
same for light polarized along the $Z$-axis and the $X$-axis,
implying that not all pairs, created along the length of either
crystal, will find any (possibly) generated pairs to interfere with
from the other crystal. Some photon-pairs will therefore be
distinguishable by temporal information.  See
Fig.~\ref{Fig:intuitive} for an illustration.  We would like to
point out that, while this chromatic two-photon dispersion effect is
reminiscent of the ``two-photon dispersion'' effects discussed in
\cite{KMWZSS95} or \cite{KFMWS04}, it does not have the same origin,
although the current effect can also be compensated for by an extra
piece of crystal. The chromatic effect comes as an disadvantage when
placing the crystals adjacent to each other, and could in principle
be avoided by an ``interferometric'' solution \cite{KKS00, FMKWS04},
in which the pump beam splits into two separate arms, impinges onto
each of the crystals, or onto a single crystal but in opposite
directions, and recombines on a beam-splitter.  Still, we believe
the current solution requires fewer optics, is easier to align, and
can be made more compact.

The previous discussion gave a limited, although intuitive,
understanding of the origin of a mixed state, but, as we will show in
the next section, a mathematical derivation will give additional
insights into how the effect of decoherence is affected by the group
velocities.

\section{The two-crystal two-photon quantum state}
In this section, we derive the output state from the two-crystal
source in terms of the frequency and polarization degrees of freedom
$\ket{\bm{\epsilon}} \otimes \ket{\bm{\chi}_{i,j}^\ph}$, where $i,j
= \{1 = ``\tV", 2 = ``\tH"\}$ denotes the polarizations. Emission
from each of the crystals, V and H, will thus be represented by
$\ket{\bm{\chi}_{11}}$ and $\ket{\bm{\chi}_{22}}$, respectively,
according to \refeqn{Eqn:onecrystalfinalstate} of the Appendix and
\reffig{Fig:sourcecoordinate}.  As just described, the vertical
light will be subject to dispersion upon its passing through the
second crystal. We will formulate this mathematically by introducing
a unitary transform acting on the states. The eigenequation which
describes the transformation $U_L$ on the state of the first
crystal, when it passes through the second crystal, is
\begin{align}
  U_L\ket{\bm{\chi}_{11}} & = e^{i (k_\ts L + k_\ti
    L)}\ket{\bm{\chi}_{11}} \nonumber\\
  & = e^{i (n_{\ts}^X \omega_{0\ts} + n_{\ti}^X \omega_{0\ti} +
    (n_{\tg,\ts}^X - n_{\tg,\ti}^X)\epsilon) L/c}
  \ket{\bm{\chi}_{11}},
\end{align}
where the length of the crystal, $L$, enters the phase term,
together with the frequency $\epsilon$.  With reference to the
Appendix \ref{App:singleCrystalState}, and
\refeqn{Eqn:onecrystalfinalstate}, we can then express the output
state of each crystal as
\begin{align}
  \ket{\Psi_{11}} &= \frac{1}{B} \int\!\! \ud \epsilon\: U_L
  U(\epsilon)\:
  \ket{\bm{\epsilon}} \otimes \ket{\bm{\chi}_{11}}, \nonumber\\
  \ket{\Psi_{22}} &= \frac{1}{B} \int\!\! \ud \epsilon\: U(\epsilon)\:
  e^{i n_{\tp}^X \omega_\tp L/c}\: \ket{\bm{\epsilon}} \otimes
  \ket{\bm{\chi}_{22}},
\end{align}
where an extra phase-term has been added to the pump field in the
second crystal due to the pump field passing through the first
crystal, $U(\epsilon)$ is the state amplitude, and $B$ is a
normalization constant. The sum of these two kets will give us the
combined two-crystal two-photon state,
\begin{align}
  \ket{\Psi^\epsilon}
  & = \ket{\Psi_{11}} + \ket{\Psi_{22}}  \nonumber\\
  & = \frac{1}{B} \int\!\! \ud \epsilon\: \Big[U_L U(\epsilon)
  \ket{\bm{\epsilon}} \otimes
  \ket{\bm{\chi}_{11}} \nonumber \\
  & \qquad +\: U(\epsilon) e^{i n_{\tp}^X \omega_\tp L/c}
  \ket{\bm{\epsilon}} \otimes
  \ket{\bm{\chi}_{22}}\Big] \nonumber\\
  & = \frac{1}{B} \int\!\! \ud \epsilon \sum \limits_{i,j = 1}^2
  c_{ij} V_{ij}(\epsilon)\: \ket{\bm{\epsilon}} \otimes
  \ket{\bm{\chi}_{ij}},
\end{align}
where we have introduced $V_{11}(\epsilon) = \frac{1}{B} U_L
U(\epsilon)$, $V_{22}(\epsilon) = \frac{1}{B} U(\epsilon) e^{i
  n_{\tp}^X \omega_\tp L/c}$, and the coefficients $c_{ij} =
1/\sqrt{2}$ for $i = j$, and $c_{ij} = 0$ for $i \not= j$,
normalized such that $ |c_{11}|^2 + |c_{22}|^2 = 1 $.

We can now form the two-photon density matrix
\begin{align}
  \bm{\rho}^{\epsilon} & = \ket{\Psi^{\epsilon}}
  \bra{\Psi^{\tilde\epsilon}} \nonumber\\
  & = \frac{1}{B^2} \iint\!\! \ud \epsilon\: \ud \tilde\epsilon\!\!
  \sum \limits_{i,j,k,l = 1}^2 \!\! c_{ij} c_{kl}^\ast V_{ij}(\epsilon)
  V_{kl}^\ast(\tilde\epsilon)\: \ket{\bm{\epsilon}}
  \bra{\tilde{\bm{\epsilon}}}\otimes \ket{\bm{\chi}_{ij}}
  \bra{\bm{\chi}_{kl}},
  \label{Eqn:densityMatrixFreq}
\end{align}
from which we would like to remove the frequency information. For
that, we need to note that we could, in principle, measure the
frequency of the photons at a resolution much smaller than the
bandwidths of the filters.  The resolution is given by a wavelength
bandwidth $\Delta\lambda_\text{res}$, which is set by the timing
information $\Delta t_\text{gate}$ of the detectors
($\Delta\lambda_\text{res} = \lambda^2/c\Delta t_\text{gate} < 8$ pm
for $\Delta t_\text{gate}> 1$ ns). Therefore, it is appropriate to
take the partial trace over the frequency mode:
\begin{align}
  \bm{\rho} & = \text{Tr}_{\epsilon}[\: \ket{\bm{\epsilon}^\prime}
  \bra{\bm{\epsilon}^\prime}\: \bm{\rho}^{\epsilon}] = \int
  \limits_{-\infty}^{\infty}\!\! \ud \bm{\epsilon}^\prime
  \bra{\bm{\epsilon}^\prime}
  \bm{\rho}^{\epsilon} \ket{\bm{\epsilon}^\prime} \nonumber\\
  & = \frac{1}{B^2} \sum \limits_{i,j,k,l = 1}^2 \! c_{ij} c_{kl}^\ast
  \int \!\!\ud \epsilon^\prime V_{ij}(\epsilon^\prime)
  V_{kl}^\ast(\epsilon^\prime)\: \ket{\bm{\chi}_{ij}}
  \bra{\bm{\chi}_{kl}}.
\end{align}

Let $\rho_{ijkl}$ denote the elements of the density matrix, of which
the only non-zero ones become
\begin{align}
  \rho_{1122} & = c_{11} c_{22}^\ast \frac{1}{B^2} \int \!\!\ud
  \epsilon^\prime U_L U(\epsilon^\prime)\: U^\ast(\epsilon^\prime)
  e^{-i n_{\tp}^X \omega_\tp L/c} \nonumber\\*
  & = \frac{1}{2} \frac{\chi_2^2\: f_1^2 E_0^2 L^2}{\hbar^2 B^2}
  \nonumber\\*
  & \quad \times\: \int\!\!\ud \epsilon^\prime\:
  |A_\ts(\epsilon^\prime)|^2 |A_\ti(\epsilon^\prime)|^2
  \nonumber\\*
  & \quad \times\: e^{-i n_{\tp}^X \omega_\tp L/c}\: e^{i (n_{\ts}^X
    \omega_{0\ts}\: +\: n_{\ti}^X \omega_{0\ti}\: +\: (n_{\tg,\ts}^X -
    n_{\tg,\ti}^X)\epsilon^\prime) L/c} \nonumber\\*
  & \quad \times\: \text{sinc}^2\left[\frac{L
      \epsilon^\prime}{2c}(n_{\tg,\ts}^Z
    - n_{\tg,\ti}^Z)\right]\: \nonumber\\*
  & = \rho_{2211}^\ast
\end{align}
and
\begin{align}
  \rho_{1111} & = \rho_{2222} = \frac{1}{2}.
\end{align}
The off-diagonal element, which describes the degree of coherence in
the entangled state, can be further simplified and identified as a
Fourier transform:
\begin{align}
  \rho_{1122} & = \frac{1}{2}\: e^{-i n_{\tp}^X \omega_\tp L/c}\: e^{i
    (n_{\ts}^X \omega_{0\ts}
    + n_{\ti}^X \omega_{0\ti}) L/c} \nonumber\\*
  & \quad \times\: \int\!\!\ud \epsilon^\prime\: g(\epsilon^\prime)\:
  e^{i \tau_X^\ph \epsilon^\prime}\:
  \text{sinc}^2(\frac{\tau_Z^\ph}{2} \epsilon^\prime),
  \label{Eqn:off-diagonalFinal}
\end{align}
where
\begin{align}
  g(\epsilon^\prime) & = \frac{\chi_2^2\:
  f_1^2 E_0^2 L^2}{\hbar^2 B^2} |A_\ts(\epsilon^\prime)|^2
  |A_\ti(\epsilon^\prime)|^2, \nonumber\\*
  \tau_X^\ph & = (n_{\tg,\ts}^X - n_{\tg,\ti}^X) L/c, \nonumber\\*
  \tau_Z^\ph & = (n_{\tg,\ts}^Z - n_{\tg,\ti}^Z) L/c.
  \label{Eqn:parameters}
\end{align}

\begin{figure}[tb!]
  \begin{center}
    \includegraphics[scale = 0.4]{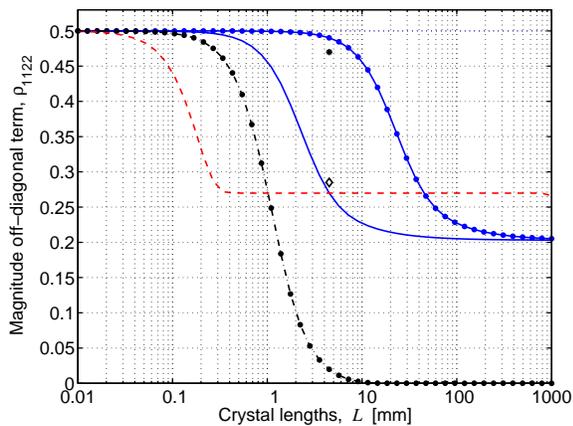}
    \caption{Color online. The off-diagonal term,
      \refeqn{Eqn:off-diagonalFinal}, of the generated density matrix
      plotted versus crystal length, which corresponds to the
      visibility of entanglement via the relation $V = 2\rho_{1122}$.
      Solid line: PPKTP-crystal with 10 nm idler filter.  Dots over
      solid line: PPKTP with 1 nm idler filter.  Dash-dotted line:
      PPLN or PPMgOLN with 10 nm idler filter.  Dashed line: PPKTP at
      optimal fiber coupling using the idler fiber's own filtering.
      Diamond: experimental value for $L = 4.5$ mm using PPKTP and
      with a 10 nm idler filter.  Solid point: experimental value for
      $L = 4.5$ mm using PPKTP and with a dispersion canceling calcite
      plate of thickness $d = 0.86$ mm.}
    \label{Fig:coherence_vs_crystal}
  \end{center}
\end{figure}

In \reffig{Fig:coherence_vs_crystal},
we have plotted the result of \refeqn{Eqn:off-diagonalFinal} versus
the length of the crystals using different crystal materials to
generate 810 and 1550 nm. We observe that the dispersion in long,
periodically poled LiNbO$_3$ (PPLN) crystal materials completely
suppresses the $\rho_{1122}$ term, and thereby the entanglement. For
PPKTP this is not the case, and if we search for $\rho_{1122}$ in
the limit of an infinitely long crystal we find that
\begin{align}
  \lim \limits_{L \to \infty} |\rho_{1122}|
  & = \left\{
    \begin{array}{ll}
      1 - \frac{\tau_X^\ph}{\tau_Z^\ph} & \text{if} \quad \tau_X^\ph <
      \tau_Z^\ph\\
      0 & \text{if} \quad \tau_X^\ph \geq \tau_Z^\ph,
    \end{array} \right.
\end{align}
which, for \mbox{PPKTP} leads to $\rho_{1122} = 0.203$, implying
still a visibility of entanglement (i.e.\ of the second-order
interference fringes) of $40.6\%$.  The different results stem from
the material-specific relation between $\tau_X^\ph$ and
$\tau_Z^\ph$. If the material is more strongly dispersive for
polarizations along the $Z$-axis than the $X$-axis, then the
off-diagonal term will be bounded below by a non-vanishing value;
otherwise, the off-diagonal term will approach zero. As expected, we
found that all numbers increase as we go closer to having degenerate
wavelength pairs. We also note that the bandwidth of the frequency
filter affects the shape of the curve; a narrower bandwidth
increases the extent of the temporal coherence and provides a
greater overlap between wave-packets, leading to an arbitrarily
increased $\rho_{1122}$. Emission that is optimally coupled into
single-mode fibers will automatically be filtered also in frequency,
since the frequency is correlated to spatial information via the
phase-matching conditions \cite{LT05}, and for long PPKTP crystals,
in such a case, the minimum value of $\rho_{1122}$ equals 0.266 ($V
= 53.2\%$).

\section{Decoherence cancellation}
\begin{figure}[b!]
  \begin{center}
    \includegraphics[scale = 0.4]{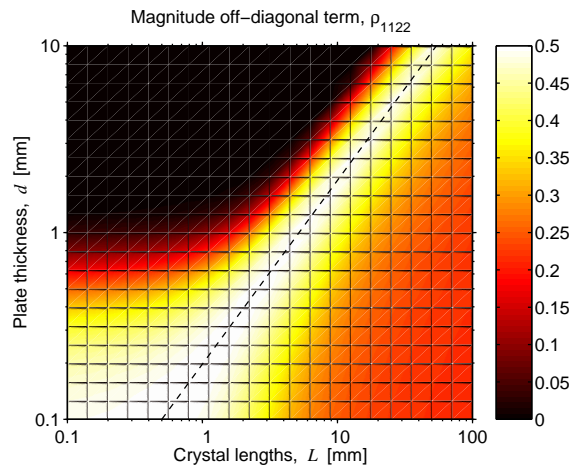}
    \caption{Color online. The off-diagonal term,
      \refeqn{Eqn:off-diagonalFinalComp}, plotted versus the crystal
      lengths, $L$, and the thickness, $d$, of a dispersion-canceling
      calcite plate, using \mbox{PPKTP} with a 10 nm idler filter. The
      dashed line represents perfect cancellation.}
    \label{Fig:compensation}
  \end{center}
\end{figure}
We will now briefly show how the pure state, $\ket{\Phi^\varphi}$ in
\refeqn{Eqn:bell-state}, can be fully regained, for generation in
long crystals, by inserting a highly birefringent crystal plate into
\text{one} of the arms. The eigenequations for each polarization
state propagating through such a crystal plate become
\begin{align}
  U_C\ket{\bm{\chi}_{1j}} & = e^{i k_\tc
    d}\ket{\bm{\chi}_{1j}} \nonumber\\*
  & = e^{i (n_{\tc}^o \omega_{0\ti} - n_{\tg,\tc}^o \epsilon) d/c}
  \ket{\bm{\chi}_{1j}}, \nonumber \\*
  U_C\ket{\bm{\chi}_{2j}} & = e^{i k_\tc
    d}\ket{\bm{\chi}_{2j}} \nonumber\\*
  & = e^{i (n_{\tc}^e \omega_{0\ti} - n_{\tg,\tc}^e \epsilon) d/c}
  \ket{\bm{\chi}_{2j}},
\end{align}
with the density matrix after the plate becoming
\begin{align}
  \bm{\rho}^{\epsilon}(d) & =
  U_C\bm{\rho}^{\epsilon}U_C^\dag.
\end{align}
Repeating Eqns. (\ref{Eqn:densityMatrixFreq}) to
(\ref{Eqn:off-diagonalFinal}), we arrive at
\begin{align}
  \rho_{1122} & = \frac{1}{2}\: e^{-i n_{\tp}^X \omega_\tp L/c}\: e^{i
    (n_{\ts}^X \omega_{0\ts}
    + n_{\ti}^X \omega_{0\ti}) L/c} \nonumber\\*
  & \quad \times\: e^{i (n_{\tc}^o \omega_{0\ti} - n_{\tc}^e
    \omega_{0\ti})
    d/c} \nonumber\\*
  & \quad \times\: \int\!\!\ud \epsilon^\prime\: g(\epsilon^\prime)\:
  e^{i (\tau_X^\ph - \kappa) \epsilon^\prime}\:
  \text{sinc}^2(\frac{\tau_Z^\ph}{2} \epsilon^\prime),
  \label{Eqn:off-diagonalFinalComp}
\end{align}
where $g(\epsilon^\prime), \tau_X^\ph$, and $\tau_Z^\ph$ is defined
by \refeqn{Eqn:parameters}, and where
\begin{align}
  \kappa & = (n_{\tg,\tc}^o - n_{\tg,\tc}^e) d/c.
\end{align}
Now, if $d$ is chosen such that $\kappa = \tau_X^\ph$, it means that
we have perfectly canceled the decoherence and retrieved a pure
state.  Hence,
\begin{align}
  \rho_{1122} = \rho_{2211}^\ast = \rho_{1111} = \rho_{2222} = \frac{1}{2}.
\end{align}
Note that, by adjusting $d$ and tilting the plate (affecting
$\varphi$) our source can prepare any arbitrary mixed state of the
kind $\bm{\rho} = V \ket{\bm{\Phi}^\varphi}\bra{\bm{\Phi}^\varphi} +
(1 - V) \bm{\rho}_m$, where $\bm{\rho}_m = \frac{1}{2}
(\ket{\bm{\chi}_{11}} \bra{\bm{\chi}_{11}} + \ket{\bm{\chi}_{22}}
\bra{\bm{\chi}_{22}})$, and $V$ is the visibility.
\reffig{Fig:compensation} shows a plot of $\rho_{1122}$ versus $L$ and
$d$.

\section{Experimental results}
\label{Sec:experimental}
\begin{figure}[b!]
  \begin{center}
    \includegraphics[scale = 0.7]{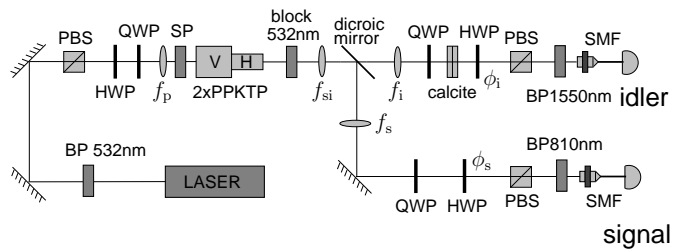}
    \caption{Experimental setup used to measure the density
      matrix. PBS: polarizing beam-splitter, HWP: half-wave plate,
      QWP: quarter-wave plate, SP: short-pass filter, BP: band-pass
      filter, SMF: single-mode fiber.}
    \label{Fig:experimentalsetup}
  \end{center}
\end{figure}

The experimental setup used when characterizing the source's output
state is shown in \reffig{Fig:experimentalsetup}.  As a pump, we use
a frequency-doubled Nd:YAG laser emitting approximately 60 mW in the
TEM$_{00}$ mode at 532 nm, which can be variably attenuated.  Its
$M_\tp^2$ value was measured to be 1.06.  After a band-pass filter
(BP532) that removes any remaining infrared light, we ``clean up''
the polarization using a polarizing beam-splitter (PBS).  The
polarization is controlled by a half-wave plate (HWP) and a
quarter-wave plate (QWP) in front of the crystal. The pump beam is
focused onto the crystal using an achromatic doublet lens ($f_\tp =
50\ \text{mm}$), which introduces a minimal amount of aberrations,
so as not to destroy the low $M^2$--value. The QWP is set to undo
any polarization elliptisation effects caused by the lens, and
fluorescence caused by the same lens is removed by a Schott-KG5
filter (SP).

\squeezetable
\begin{table*}[tb!]
  \caption{\label{Tab:rates} Three runs at different alignments and
    pump powers, showing the coupling efficiencies, photon rates in
    fibers, conversion efficiency, and the production rate of the
    system.}
  \begin{ruledtabular}
    \begin{tabular}{dccccccccccc}
      \multicolumn{1}{c}{$P_\tp$ [mW]} & $\gamma_\ts$ & $\gamma_\ti$ &
      $\gamma_\tc$ & $\mu_{\ti|\ts}$ & $\sigma$ & $R_\ts$ [s$^{-1}$] &
      $R_\ti$ [s$^{-1}$] & $R_\tp$ [s$^{-1}$] & $R_\tc$ [s$^{-1}$] &
      $\beta$ & $R_\tc^\text{prod}$ [\prodrateunit{\!}]
      \rule[-2ex]{0pt}{3ex}\\
      \hline
      \rule{0pt}{3ex}
      60 & 0.32 & 0.79 & 0.11 & 0.12 & 0.34 & $2.32 \times 10^6$ &
      $2.39 \times 10^6$ & $8.61 \times 10^6$ & $274
      \times 10^3$  & $5\times10^{-11}$ & $5.0
      \times 10^3$\\
      4.5 & 0.32 & 0.56 & 0.10 & 0.11 & 0.32 & $167 \times 10^3$ &
      $121 \times 10^3$ & $617 \times 10^3$ & $19
      \times 10^3$  & $5\times10^{-11}$ & $4.6
      \times 10^3$\\
      0.54 & 0.46 & 0.38 & 0.22 & 0.27 & 0.57 & $100 \times 10^3$ &
      $195 \times 10^3$ & $450 \times 10^3$ & $27
      \times 10^3$  & $3\times10^{-10}$ & $55
      \times 10^3$\\
    \end{tabular}
  \end{ruledtabular}
\end{table*}

The next components are the two PPKTP crystals, which are heated in
an oven to a temperature $T \approx 100^\circ$.  After the crystals,
we block the pump light with a 532 nm band-stop filter, and the
signal and idler emission is focused by achromatic doublet lenses.
To separate the 810 nm and 1550 nm emission, we use a dichroic
mirror made for a $45^\circ$ angle of incidence.  The first lens
($f_{\ts\ti} = 30\ \text{mm}$) is common to both signal and idler,
and its task is to refocus the beams somewhere near the dichroic
mirror.  The next two lenses ($f_\ts = 60\ \text{mm}$ and $f_\ti =
40\ \text{mm}$) collimate each beam, which are then focused into the
fiber tips (with the mode field diameters being $\text{MFD}_{810} =
5.5\ \mu\text{m}$ and $\text{MFD}_{1550} = 10.4\ \mu\text{m}$) using
aspherical lenses with $f = 11 \text{\ mm}$. Next, we use
quarter-wave plates (QWP), half-wave plates (HWP), and polarizing
beam-splitters (PBS) in each arm to analyze the state. In the idler
arm, we also place the tiltable cancellation plate, which is made of
calcite. In front of the fiber couplers, we have first
Schott-RG715/RG1000 filters to block any remaining pump light, and
then interference filters (BP) of 2 nm and 10 nm bandwidth at the
810 nm and 1550 nm side respectively. The detectors used are a
Si-based APD (PerkinElmer SPCM-AQR-14) for 810 nm with a quantum
efficiency $\eta_\ts = 60\%$ and a homemade InGaAs-APD (Epitaxx)
module for 1550 nm with $\eta_\ti = 18\%$, gated with 5 ns pulses.
To avoid afterpulsing effects, the InGaAs-APD is used together with
a hold-off circuit ($10\ \mu$s) for all of the measurements. The
pulses were generated using a digital delay generator (DG535) from
SRS, with a maximal repetition rate of 1 MHz, and a trigger dead
time of $1\ \mu$s.

\begin{figure}[tb!]
  \begin{center}
    \includegraphics[scale = 0.4]{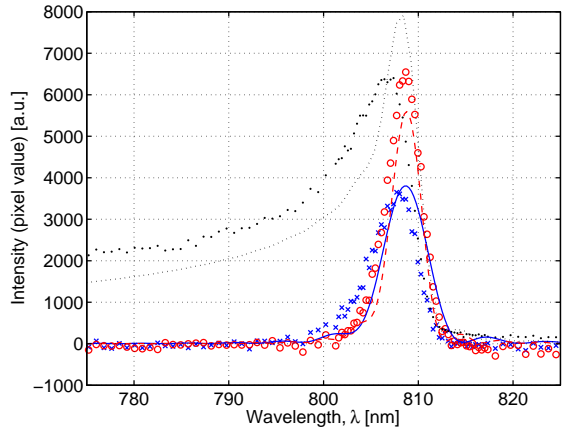}
    \caption{Color online. Spectrogram of signal emission inside
      single-mode fibers without interference filters. The crosses
      ($\Delta\lambda_\ts = 6$ nm) and circles ($\Delta\lambda_\ts =
      4$ nm) represent experimental data for the H and the V crystal,
      respectively, to be compared to theoretical predictions for a 2
      nm (solid line) and a 3 nm (dashed line) long crystal. Also
      shown is the downconversion spectrum in free space,
      experimentally (dots) and in theory (dotted line), demonstrating
      the fiber's own filtering.}
    \label{Fig:spectrum_signal}
  \end{center}
\end{figure}

We have used a spectrograph (SpectraPro 500i, ARC) to measure the
bandwidth of the signal emission using a single-mode fiber without
any filter; see \reffig{Fig:spectrum_signal}.
The bandwidth was found to be 4 nm for the V-crystal and 6 nm for
the H-crystal.  The results in \cite{LT05} suggests that the
effective lengths of the crystals being poled must then be 3 mm and
2 mm, respectively, but also that the 2 mm crystal should give
$\approx 55\%$ of the photon-rate of the 3 mm one.  Experimental
agreement is good, as we saw the H-crystal giving half the rate of
the V-crystal. When measuring, we refocused the fiber coupling for
each crystal to find maximum counts, while keeping the pump
polarization exactly at $45^\circ$. As described in connection to
\reffig{Fig:venn_diagram}, the best tradeoff when collecting from
both crystals simultaneously is to set the focus of the pump mode
and the fiber-matched modes at the intersecting faces.
Experimentally, however, in order to produce as pure a Bell-state as
possible, we needed to balance the rate of each crystal, which we
did by shifting the fiber-matched focus a bit closer to the
H-crystal and by turning the pump-polarization slightly towards H.
(The focus point was moved by turning the focusing knob on the fiber
coupler.) In this way we allowed lower coupling efficiencies than
the maximum attainable. The focusing conditions achieved with
available lenses were, $\xi_\tp = 2.1$ for the pump mode, $\xi_\ts =
3.2$ for the signal's fiber-matched mode, and $\xi_\ti = 2.5$ for
the idler's.

With this configuration, we obtained the results showed in
\reftab{Tab:rates}.  In each column of the table, $\gamma_\ts$
represents the signal's single coupling efficiency, $\gamma_\ti$ the
idler's, $\gamma_c$ the pair coupling efficiency, $\mu_{\ti|\ts}$
the conditional coincidence, and $\sigma$ the correlation
efficiency, which corresponds directly to $\mu_{\ti|\ts}$ but
includes a compensation for the $35\%$ transmission of the 1550 nm
filter and the $85\%$ transmission of the 810 nm filter. The singles
photon-rate in the signal fiber, $R_\ts$, and the idler $R_\ti$,
were both derived from detected raw counts. The total generated rate
$R_\tp$ of pairs before fiber coupling was estimated from detected
counts using a multimode fiber. The pair rate in the fibers,
$R_\tc$, was deduced from the above efficiencies and the detected
raw coincidence rate, with accidental counts subtracted by assuming
that $R_\ti$ originates from a Poissonian distribution at random
gating \cite{TL05}.  The conversion efficiency $\beta$ is the
fraction of pump photons converted into signal and idler pairs,
leading to a pair production rate $R_\tc^\text{prod}$, which equals
\prodrateunit{$5
  \times 10^3$} at the pump power $P_\tp = 60\ \text{mW}$ and with the
idler detector gated at 585 kHz. (The production rate is the pair
rate normalized to the wavelength bandwidth in THz and the pump
power in mW.) The second row of \reftab{Tab:rates} shows similar
results for a lower pump power and an idler gate rate of 91 kHz.  We
also took measurements without any interference filter at the idler
side (but still with a 2 nm filter at the signal), with the results
shown in the third row of \reftab{Tab:rates}, for $P_\tp = 540\
\mu\text{W}$ and with a gate-rate of 57 kHz. The results are
improved, not because of the lower power, but because of a
simultaneous optimization of the arms in order to maximize
$\gamma_\tc$. The table shows how $\gamma_\ti$ decreases in the
process.  The correlation efficiency $\sigma$ now includes the
estimated transmission-loss of the optics at the idler side, and a
correction factor for the unequal filtering between signal and idler
(the idler fiber itself provides a frequency filtering of
$\Delta\lambda_\ti = 14.7$ nm). The best conditional coincidence is
$\mu_{\ti|\ts} = 0.27$, and the conversion efficiency, $\beta = 3
\times 10^{-10}$, was possibly improved by aligning to a more
homogeneously poled area of the crystals. We believe that the pair
production rate, \prodrateunit{$55 \times 10^3$}, is one of the
highest yet reported for polarization entangled photon pairs
generated in crystals and launched into single-mode fibers.
Frequency filtering at a narrow bandwidth of 50 GHz would imply $3
\times 10^3$ s$^{-1}$mW$^{-1}$ of pairs in the fibers.  Besides, for
narrow filtering, the photon flux has been shown \cite{LT05} to be
$\propto L\sqrt{L}$, and so, by using longer crystals ($L = 50$ mm)
we could still reach 20 s$^{-1}$mW$^{-1}$ at a 10 MHz bandwidth,
which is the bandwidth regime of e.g.\ Rb-atom based quantum
memories.  As a comparison, we have derived numbers using data
available for some other experiments, among which the best include
Fiorentino \etal \cite{FMKWS04}, who seem to have \prodrateunit{$22
\times 10^3$} of pairs being generated by two 10 mm long crystals
into free-space; König \etal \cite{KMWA05}, who claim to have
\prodrateunit{$300 \times
  10^3$} pairs from two 20 mm long crystals into fibers; and Li \etal
\cite{LVSK05}, who seem to have an exceptional value of
\prodrateunit{$4.3 \times
  10^6$} pairs generated directly inside a non-linear fiber.

\begin{figure}[tb!]
  \begin{center}
    \includegraphics[scale = 0.4]{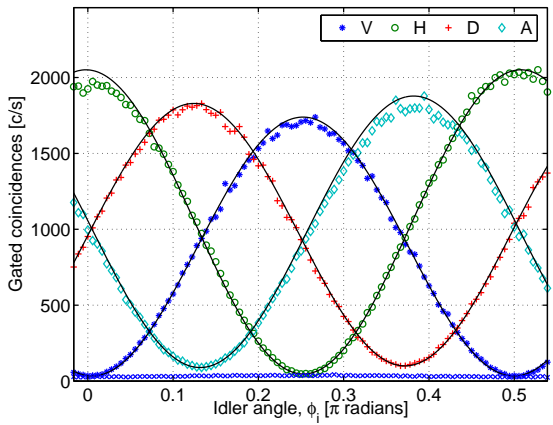}
    \caption{Color online. The plot shows the raw visibility curves
      obtained at a pump power $P_\tp = 4.5$ mW, gate-rate 91 kHz,
      without subtraction of background counts (including accidentals)
      shown at the bottom.  Each curve correspond to a different
      polarization setting of the signal arm, $\phi_\ts^\tV =
      -3\pi/16$, $\phi_\ts^\tH = \pi/16$, $\phi_\ts^\tD = -\pi/16$,
      and $\phi_\ts^\tA = 3\pi/16$, as indicated the inset.  Fitting
      of the collected data gives $V_\tH = 99.6 \pm 0.2\%$, $V_\tV =
      99.2 \pm 0.2\%$, $V_\tD = 92.7 \pm 0.2\%$, and $V_\tA = 94.2 \pm
      0.2\%$. When background counts were not subtracted, we obtained
      the visibilities $V_\tH = 95.6 \pm 0.2\%$, $V_\tV = 96.2 \pm
      0.2\%$, $V_\tD = 89.6 \pm 0.2\%$, and $V_\tA = 90.9 \pm 0.2\%$.}
    \label{Fig:visibilities}
  \end{center}
\end{figure}

\reffig{Fig:visibilities} shows the visibility curves obtained, with
and without subtraction of background counts, including false
coincidences. Note that the number of ``accidental'' coincidences
increases with the pump power, as the probability of more than a
single pair to arrive within the gate-time of the detector
increases, as shown in section II.

We have also measured a violation of the CHSH-inequality
\cite{CHSH69} by taking measurements of the coincidence-rate
functions
\begin{align}
  R_{i,j} = \frac{1}{2}[1 + i j V_{i,j} \cos(4\phi_\ts + 4\phi_\ti)],
\end{align}
where $i, j = \pm 1$ denotes the four combinations of measurable
output-arms of the two PBS:s in the signal and idler, $V_{i,j}$ is
the corresponding visibility, and $\phi_\ts$, $\phi_\ti$, are the
angles of the HWPs. The correlation function becomes
\begin{align}
  E(\phi_\ts, \phi_\ti) &= \frac{R_{1,1} - R_{1,-1} - R_{-1,1} +
    R_{-1,-1}}{R_{1,1} + R_{1,-1} + R_{-1,1} + R_{-1,-1}} \nonumber \\
  &= V \cos(4\phi_\ts + 4\phi_\ti),
  \label{Eqn:correlation-fcn}
\end{align}
where $V = V_{1,1}$, assuming fully equal rate functions, so that we
can rely on measurements taken at only one of the output arms.
Entanglement is present iff the CHSH-inequality is violated,
\begin{align}
  S = E(\phi_\ts^1, \phi_\ti^1) + E(\phi_\ts^1, \phi_\ti^2) +
  |E(\phi_\ts^2, \phi_\ti^1) - E(\phi_\ts^2, \phi_\ti^2)| \leq 2,
  \label{Eqn:CHSH-inequality}
\end{align}
where the correlation function is to be measured at the following
pair of angles: $\phi_\ts^1 = -\pi/16, \phi_\ti^1 = 0$ and
$\phi_\ts^2 = \pi/16, \phi_\ti^2 = \pi/8$. The parameter $S$ can
reach the maximum value of $2\sqrt{2}$, corresponding to $100\%$
visibility, and it is well known that the average visibility needs
to be $> 71\%$ to violate the inequality, if the state is subject to
equal decoherence in all bases. In our case, the state decoheres in
the H/V-basis, while maintaining nearly perfect visibility for the H
and V settings. Let $\phi_\ts^1$ represent the H/V-basis, and
$\phi_\ts^2$ the D/A-basis. Furthermore, let $V_{\tH, \tV}$ and
$V_{\tD, \tA}$ represent the visibilities in each respective basis.
We get
\begin{align}
  S &= V_{\tH,\tV}\cos(-\frac{\pi}{4} + 0) +
  V_{\tH,\tV}\cos(-\frac{\pi}{4} + \frac{\pi}{2}) + \nonumber \\
  & \quad \left|V_{\tD,\tA}\cos(\frac{\pi}{4} + 0)
  - V_{\tD, \tA}\cos(\frac{\pi}{4} + \frac{\pi}{2})\right| \nonumber \\
  &= \sqrt{2}(V_{\tH,\tV} + V_{\tD,\tA}),
  \label{Eqn:S_from_visibilities}
\end{align}
which shows that, for $V_{\tH,\tV} = 100\%$, the requirement is
$V_{\tD,\tA} > 41\%$ for a violation of
\refeqn{Eqn:CHSH-inequality}.


A direct measurement of $S$ at the above angles yields $S = 2.679 \pm
0.004$ at a pump power $P_\tp = 60$ mW, after subtraction of
accidental counts (gate-rate was 585 kHz). The CHSH-inequality was
violated by 177 standard deviations in 10 s, or $56\sigma_S^\ph\ 
\text{s}^{-1/2}$. To our knowledge, this is one of the highest
reported to date; only Kurtsiefer \etal \cite{KOW01a} exceeds this
rate, with $148\sigma_S^\ph\ \text{s}^{-1/2}$.  Other examples of good
results can be found in \cite{KWWAE99} ($50\sigma_S^\ph\ 
\text{s}^{-1/2}$) and in \cite{FMKWS04} ($38\sigma_S^\ph\ 
\text{s}^{-1/2}$).  We have re-derived these numbers using available
data, in the hopes of having created directly comparable normalized
numbers. The derivation was made as follows.  Assuming no fluctuation
of the rate other than that originating from Poissonian-distributed
single-photon detections, the standard deviation of the coincidence
rate $R_{i,j}$ becomes $\sigma_R^\ph =
{\sqrt{R_\text{max}/2}}/{\sqrt{T_R}}$, where $R_\text{max}$ is the
peak coincidence rate, $\sqrt{R_\text{max}/2}$ is the standard
deviation of the average photon-rate, $T_R$ is the integration-time in
seconds, and the central limit theorem is used to sum over time.
According to \refeqn{Eqn:correlation-fcn}, the standard deviation of
the correlation function becomes $\sigma_E^\ph = \sqrt{4} \sigma_R^\ph
/ 2 R_\text{max}$, and by \refeqn{Eqn:CHSH-inequality} we have
$\sigma_S^\ph = \sqrt{4} \sigma_E^\ph = 2/\sqrt{2 R_\text{max} T_R}$,
such that $S = S_m \pm \sigma_S^\ph$, where $S_m$ is the measured
value over $T_R$ seconds.  Thus, the normalized ``speed of CHSH
violation'' becomes
\begin{align}
  x = \frac{S_m - 2}{\sigma_S^\ph \sqrt{T_R}} = \frac{(S_m - 2)\sqrt{2
      R_\text{max}}}{2} \quad [\text{s}^{-1/2}],
\end{align}
which only depends on the maximum rate and the measured value of
$S$. If the accidental counts are not subtracted from the
coincidence counts, we instead measure the value $S = 2.6283 \pm
0.0102$ ($P_\tp = 4.5$ mW), with the CHSH-inequality being violated
by $19\sigma_S^\ph\ \text{s}^{-1/2}$, showing that we truly have a
high degree of entanglement launched into the fibers. This is
important in entanglement-based quantum key distribution (QKD)
systems that do not allow a subtraction of the background. Rather,
any accidentals will increase the quantum bit error rate (QBER) and
reduce the final bit rate, equivalently degrading the system
performance.

\begin{figure}[tb!]
  \begin{center}
    \includegraphics[scale = 1]{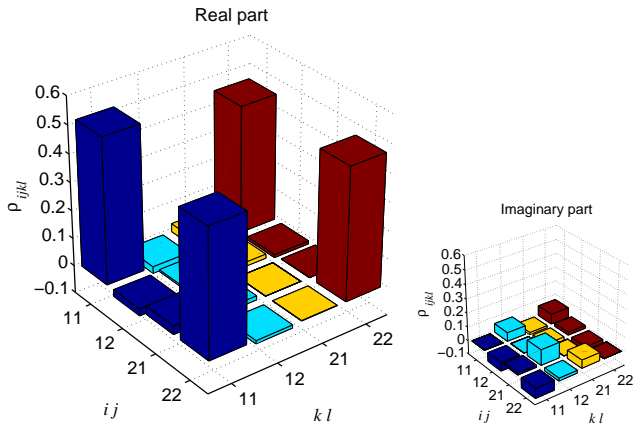}
    \caption{Color online. Experimentally determined density matrix,
      $\bm{\rho}_\text{exp}$, (real and imaginary parts) obtained by
      quantum state tomography on the generated polarization entangled
      state (1 = ``V'' and 2 = ``H''). (Pump power, $P_\tp = 4.5$
      mW).}
    \label{Fig:densityexp}
  \end{center}
\end{figure}

Following Ref.\ \cite{JKMW01}, we have made a complete tomography of the
state, with the resulting density matrix becoming
\begin{multline}
  \bm{\rho}_\text{exp} =
  \left[
    \begin{smallmatrix}
      0.5197 & -0.0237 & 0.0300 & 0.4573 \\
      -0.0237 & 0.0069 & 0.0146 & -0.0114 \\
      0.0300 & 0.0146 & 0 & 0.0010 \\
      0.4573 & -0.0114 & 0.0010 & 0.4734
    \end{smallmatrix}
  \right]\\
  + i\left[
    \begin{smallmatrix}
      0 & 0.0628 & - 0.0150 & 0.0720 \\
      - 0.0628 & 0 & - 0.1107 & 0.0206 \\
      0.0150 & 0.1107 & 0 & -0.0581 \\
      - 0.0720 & - 0.0206 & 0.0581 & 0
    \end{smallmatrix}
  \right],
\end{multline}
which is also plotted in \reffig{Fig:densityexp}. Recall that the
off-diagonal element, $\rho_{1122} = 0.457$, corresponds approximately
to the visibility in the D/A-basis, $V \approx 2\rho_{1122} = 0.915$,
which is indeed close to the measured visibilities. When applying the
density matrix to Wootters's entanglement of formation measure
\cite{Wootters98}, we get the value $E = 0.56$. The entanglement of
formation equals unity for a pure Bell-state, as do the fidelity, $F =
\bra{\bm{\Phi}^\varphi} \bm{\rho}_\text{exp} \ket{\bm{\Phi}^\varphi}$,
which is found to be 0.95 for the generated state.

\section{Future directions: A hybrid-coded entanglement source}

\begin{figure}[tb!]
    \begin{center}
    \includegraphics[scale = 0.7]{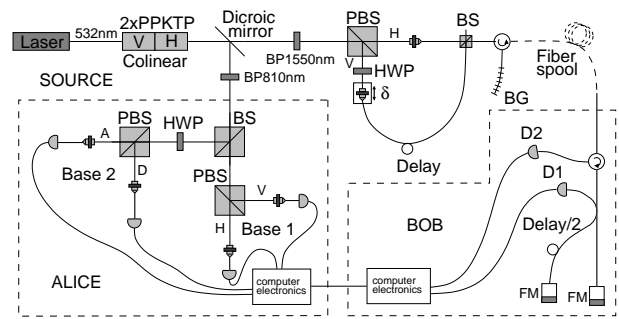}
    \caption{Scheme to create hybrid-coded entanglement. BS:
      beam splitter, PBS: polarizing beam splitter, HWP: half-wave
      plate, BP: band-pass filter, BG: Bragg grating, FM: Faraday
      mirror.}
    \label{Fig:hybrid_setup}
    \end{center}
\end{figure}

In order to motivate the usefulness of the source, we provide in
\reffig{Fig:hybrid_setup} a complete setup for quantum communication
(e.g.\ QKD). The scheme, which is under implementation, uses long
crystals ($2\times50$ mm) in order to achieve a bandwidth of $<80$
GHz, which means higher production rates and less dispersion in
combination with a telecom Bragg grating as dispersion compensator.
For long crystals, the optimal focusing is weaker, which leads to a
more compact source with fewer collimating lenses placed at closer
distances to each other.  Furthermore, improvement of the
conditional coincidences as well as the size of the source can be
achieved by minimizing the number of components, each of which
contribute to loss.

In Bob's arm, the polarization information is converted into time
information in order to avoid the polarization dispersion in
standard telecom fibers.  (For a thorough review on photonic qubits,
please refer to \cite{TW01}.) A polarizing beam splitter sits in an
unbalanced Mach-Zehnder interferometer, directing vertical photons
into the long arm and horizontal into the short. The vertical
photons are rotated to horizontal before the photons in both arms
are recombined on a fiber-based beam-splitter and sent to a Bragg
grating.  The result is a time encoded qubit with all polarization
information erased. (To our knowledge, there is no way to erase this
information passively without having to accept 50\% losses in the
unused arm of the beam-splitter, which is an disadvantage, but could
also be turned to an advantage by introducing a third party
Charlie.)  The resulting state becomes $ \ket{\Phi^\prime} = 1 /
\sqrt{2} \left(\ket{\tV}_\ts
  \ket{\text{L}}_\ti + e^{i\varphi} \ket{\tH}_\ts
  \ket{\text{S}}_\ti\right)$, where L denotes the long arm and S the
short arm.
On Bob's analyzer side, there is an unbalanced all-fiber Michelson
interferometer with a single beam-splitter to decode the qubits. The
interferometer uses Faraday mirrors, which reflect the light in such
a way that the polarization is exactly orthogonal when the photons
arrive a second time at the beam-splitter to interfere, and thereby
avoids the need for polarization controllers \cite{TBZG98}. The
phase information of the qubit defines a complementary basis to
time, and for that information to remain, the path length difference
between the short and the long arm needs to be exactly matched to
that of the preparing interferometer, requiring both interferometers
to be temperature stabilized. However, longer coherence length of
the emitted photons (an effect of narrow bandwidth) will effectively
relax these requirements.  One advantage of the above solution is
that the preparing interferometer has translatable fiber couplers
inside the interferometer, which simplifies their mutual alignment.
Also, we avoid the possibly difficult alignment of three
interferometers, as Alice adheres to polarization coding. Another
important condition for the qubits to remain coherent is that the
delay between two consecutive pulses is short enough ($\approx 5$
ns) that they experience the same phase shift due to vibrations and
temperature fluctuations when traveling over the fiber.  On Alice's
side, the analyzer realizes a standard polarization decoder. Note
that the H/V or D/A-basis is randomly chosen by the first
beam-splitter, just as at Bob's side, which implies that there is no
need for any active devices. Note also that there exists the
possibility to delay the outputs of each detector arm on Alice side
and combine into different time-slots for detection with a single
detector, instead of four, which may reduce the need for space.

\section{Summary}

In this article, we have presented work on a two-crystal source that
uses PPKTP for the production of polarization-entangled photon-pairs
in a single spatial mode, leading to efficient fiber coupling. The
source is suitable for schemes that combine polarization and time
coding.  We have shown how distinguishability between photon-pairs
is introduced for this type of colinear source, due to a special
kind of chromatic two-photon dispersion. We have derived and
analyzed the output state of SPDC for this case, with the goal to
cancel the decoherence and regain a pure state using an extra piece
of birefringent crystal.  We have determined the quality of
entanglement for the reported setup using various measures,
including the method of quantum state tomography, and we draw the
conclusion that this is one of the brightest sources available for
polarization entanglement in terms of Bell-inequality violation and
production rates.

\begin{acknowledgments}
  The authors would like to thank A. Karlsson and G. Björk for their
  valuable comments and suggestions throughout the work, M.  Andersson
  and J. Tidström for useful discussions, A.  Fragemann, C.  Canalias,
  and F.  Laurell for providing us with crystals, and J.  Waldebäck
  for his help with electronics. Financial support is gratefully
  acknowledged from the European Commission through the integrated
  project \mbox{SECOQC} (Contract No.  IST-2003-506813), and from the
  Swedish Foundation for Strategic Research (\mbox{SSF}).
\end{acknowledgments}

\appendix*
\section{The two-photon frequency and polarization quantum state}
\label{App:singleCrystalState} In this Appendix, we derive the
quantum state of a single crystal in terms of frequency and
polarization degrees of freedom, using the interaction picture of
SPDC \cite{Klyshko88}.

The evolution of the number state vector is given by
\begin{align}
  \ket{\psi} &= \exp\left[\frac{1}{i\hbar} \int \limits_{T}^{t_0 +
      T}\!\!  \ud t
    \hat H(t)\right]\ket{\psi_{00}}\nonumber\\*
  & \approx \left(\openone + \frac{1}{i\hbar} \int \limits_{T}^{t_0 +
      T}\!\! \ud t \hat H(t)\right)\ket{\psi_{00}},
  \label{Eqn:evolution}
\end{align}
where $\ket{\psi_{00}}$ is the number state at time $t_0$ and $\hat
H(t)$ is the interaction Hamiltonian,
\begin{align}
  \hat H(t) = \int \limits_{-L/2}^{L/2}\!\!\!\! \ud z\!\! \int
  \limits_{-\infty}^{\infty}\!\!\! \ud y\!\! \int
  \limits_{-\infty}^{\infty}\!\!\! \ud x\: \chi^{(2)} \hat
  E_\tp^{(+)}\hat E_\ts^{(-)}\hat E_\ti^{(-)} + \mathrm{H.c.},
\end{align}
displayed in a Cartesian coordinate system, $ \bm{r} = x\bm{e}_x +
y\bm{e}_y + z\bm{e}_z$. There are three interacting fields in the
crystal's volume, ignoring all higher-order terms ($n \geq 3$) of
the non-linearity $\chi^{(n)}$. All three fields have the same
polarization ($ZZZ$):
\begin{align}
  &  E_\tp^{(+)} = E_0\: e^{-i(k_{0\tp} \bm{s}_\tp \cdot \bm{r} -
  \omega_\tp t + \phi_\tp)} \\*
  & \hat E_\ts^{(-)} =\! \int\!\! \ud \phi_\ts \int\!\! \ud \omega_\ts
  A_\ts(\omega_\ts) \sum \limits_{\bm{s}_\ts}
  e^{i(k_\ts \bm{s}_\ts \cdot \bm{r} - \omega_\ts t + \phi_\ts)}
  \hat{a}_\ts^\dagger(\omega_\ts, \bm{s}_\ts)\!\!\! \\*
  & \hat E_\ti^{(-)} =\! \int\!\! \ud \phi_\ti \int\!\! \ud \omega_\ti
  A_\ti(\omega_\ti) \sum \limits_{\bm{s}_\ti}
  e^{i(k_\ti \bm{s}_\ti \cdot \bm{r} - \omega_\ti t +
  \phi_\ti)}\hat{a}_\ti^\dagger(\omega_\ti, \bm{s}_\ti),
\end{align}
where the pump field is classical and monochromatic so that we can
replace $\hat E_\tp^{(+)} $ by $ E_\tp^{(+)}$. The plus sign denotes
conjugation, i.e annihilation (+) or creation (-) of the state. We
have also introduced the notation $\bm{k} = k \bm{s}$, where $\bm{s}
= p \bm{e}_x + q \bm{e}_y + m \bm{e}_z$, is the unit length vector
of $\bm{k}$ with components in each of the three dimensions
\cite{MW95}, as defined by the coordinate system in
Fig.~\ref{Fig:sourcecoordinate}. The pump field is a plane wave
propagating in the $z$-direction, $\bm{s}_\tp = \bm{e}_z $. For
signal and idler, we sum over both frequency and angular modes,
where $\hat{a}(\omega, \bm{s})$ is the field operator, and
$A(\omega)$ is the frequency amplitude of a Gaussian-shaped detector
filter having the bandwidth $\Delta\lambda$ ({\small FWHM}) and
center wavelength $\lambda_c^\ph$ (all wavelengths in vacuum). Via
the relation $\omega = 2\pi c n_{\!  \scriptscriptstyle \lambda} /
\lambda$, its form is given by
\begin{align}
  A(\omega; \lambda) = e^{-2\log(2)(\lambda - \lambda_c^\ph)^2 /
    \Delta\lambda^2 }.
\end{align}
Each signal and idler photon is created with a random phase,
$\phi_\ts$ and $\phi_\ti$, respectively, which we need to sum over.
The phase of the pump, $\phi_\tp$, is constant and arbitrary.

For periodically poled materials, the spatial variation of the
nonlinear index $\chi^{(2)}$ has sharp boundaries, but we will
simplify and make a sinusoidal approximation using the first term of
an Fourier-series expansion of $\chi^{(2)}$:
\begin{align}
  \chi^{(2)} = \chi_2^\ph \sum_{m = 0}^\infty f_m e^{-i m \bm{K} \cdot
    \bm{r}} \approx \chi_2^\ph\: f_1 e^{-i \bm{K}\cdot \bm{r}},
  \label{Eqn:chi2}
\end{align}
where $\bm{K} = K \bm{e}_z = {2 \pi / \Lambda}\: \bm{e}_z$ and
$\Lambda$ is the grating period.

The Hamiltonian now takes the form
\begin{align}
  \hat H(t) & = \chi_2^\ph\: f_1 E_0 \int\!\!\ud \phi_\ts \int\!\!\ud
  \phi_\ti \int\!\!\ud \omega_\ts \int\!\!\ud
  \omega_\ti\: \nonumber\\
  & \quad \times\: A_\ts(\omega_\ts) A_\ti(\omega_\ti) \nonumber\\
  & \quad \times\: \sum \limits_{\bm{s}_\ts} \sum \limits_{\bm{s}_\ti}
  \hat{a}_\ts^\dag(\omega_\ts, \bm{s}_\ts)
  \hat{a}_\ti^\dag(\omega_\ti, \bm{s}_\ti) \nonumber\\*
  & \quad \times\: \int \limits_{-L/2}^{L/2}\!\!\!\! \ud
  z\!\! \int \limits_{-\infty}^{\infty}\!\!\! \ud y\!\! \int
  \limits_{-\infty}^{\infty}\!\!\! \ud x \nonumber \\*
  & \quad \times\: e^{-i[\Delta \bm{k}\cdot (x \bm{e}_x + y \bm{e}_y + z
    \bm{e}_z)\: -\: (\omega_\ts + \omega_\ti - \omega_\tp)t\: +\: \phi_\ts
    + \phi_\ti -
    \phi_\tp]}\nonumber\\
  & \quad +\: \mathrm{H.c.},
  \label{Eqn:hamiltonian}
\end{align}
where the mismatch vector is
\begin{align}
  \Delta \bm{k} &=k_\ts \bm{s}_\ts + k_\ti \bm{s}_\ti - k_{0\tp}
  \bm{s}_\tp + \bm{K}\nonumber\\
  & = \Delta k_x \bm{e}_x + \Delta k_y \bm{e}_y + \Delta k_z \bm{e}_z.
  \label{Eqn:mismatchvector}
\end{align}

Following \refeqn{Eqn:evolution}, we now let the Hamiltonian undergo
time evolution. The mismatch vector is also divided up into
its x, y, and z components using \refeqn{Eqn:mismatchvector}. Hence,\\

\lefteqn{\frac{1}{i\hbar} \int\!\! \ud t \hat H(t) = \nonumber}
\nopagebreak \vspace{-3mm}
\begin{align}
  & \quad \chi_2^\ph\: f_1 E_0 \int\!\!\ud \omega_\ts \int\!\!\ud
  \omega_\ti\: A_\ts(\omega_\ts) A_\ti(\omega_\ti) \nonumber\\
  & \quad \times\: \sum \limits_{\bm{s}_\ts} \sum \limits_{\bm{s}_\ti}
  \hat{a}_\ts^\dag(\omega_\ts, \bm{s}_\ts)
  \hat{a}_\ti^\dag(\omega_\ti, \bm{s}_\ti) \nonumber\\*
  & \quad \times\: \int \limits_{-L/2}^{L/2}\!\!\!\! \ud z\!\! \int
  \limits_{-\infty}^{\infty}\!\!\! \ud y\!\! \int
  \limits_{-\infty}^{\infty}\!\!\! \ud x\: e^{-i[\Delta k_x x + \Delta
    k_y y + \Delta k_z z]}\: \nonumber \\*
  & \quad \times\: \frac{1}{i\hbar} \iint \limits_0^{\quad 2\pi}\! \ud
  \phi_\ts \ud \phi_\ti \int \limits_0^T\! \ud t\: e^{-i[(\omega_\ts +
    \omega_\ti - \omega_\tp)t\: +\: \phi_\ts + \phi_\ti -
    \phi_\tp]}\nonumber\\
  & \quad -\: \mathrm{H.c.}.
  \label{Eqn:hamiltoniantime1}
\end{align}

The integration over the interaction volume, $\ud x$, $\ud y$, and
$\ud z$, can now be easily carried out. There are three spatial
integrals, of which two are the Fourier transforms of unity ($\ud x$
and $\ud y$) and one is the transform of a box function ($\ud z$).
The transforms turn into two $\delta$ functions and a sinc function,
respectively. The time integral also turns into a $\delta$ function
of the three frequencies $\omega_\ts$, $\omega_\ti$, and
$\omega_\tp$. This is because we assume a monochromatic pump beam
with infinite coherence length, which effectively leads to an
infinite interaction time, ${T \rightarrow \infty}$, even for short
crystals. Motivated by the rotational symmetry of the emitted modes,
we also change to a spherical coordinate system (see
Fig.~\ref{Fig:sourcecoordinate}), by replacing the summation over
$\bm{s}$ with integrals over $\theta_\ts, \theta_\ti, \varphi_\ts$
and $\varphi_\ti$. Furthermore, the only non-zero solution for the
integration over the random phases, $\phi_\ts$ and $\phi_\ti$, is
for the phases to add up to a constant, yielding the relation
${\phi_\ts +
  \phi_\ti = \phi_\tp + C}$. If we let $C = 0$ for simplicity, and
drop some constants resulting from the integrations, we are led to
\begin{align}
  \frac{1}{i\hbar} \int\!\! \ud t \hat H(t) & = \frac{1}{i\hbar}
  \chi_2^\ph\: f_1 E_0 \int\!\!\ud \omega_\ts \int\!\!\ud
  \omega_\ti\: A_\ts(\omega_\ts) A_\ti(\omega_\ti) \nonumber \\
  & \quad \times\: \int \limits_0^{\pi/2}\!\! \sin \theta_\ts\: \ud \theta_\ts
  \int \limits_0^{\pi/2}\!\! \sin \theta_\ti\: \ud \theta_\ti \int
  \limits_0^{2\pi}\!\! \ud \varphi_\ts \int
  \limits_0^{2\pi}\!\! \ud \varphi_\ti\: \nonumber \\
  & \quad \times\: \hat{a}_\ts^\dag(\omega_\ts, \theta_\ts, \varphi_\ts)\:
  \hat{a}_\ti^\dag(\omega_\ti, \theta_\ti, \varphi_\ti)\:
  \delta(\omega_\ts +
  \omega_\ti - \omega_\tp) \nonumber \\
  & \quad \times\: \delta(\Delta k_x)\: \delta(\Delta k_y)\: L\:
  \text{sinc}\left[\frac{L}{2}\Delta k_z\right] \nonumber \\*
  & \quad -\: \mathrm{H.c.}.
  \label{Eqn:hamiltoniantime2}
\end{align}

At this stage, we observe that $k_\ts$ and $k_\ti$ each depend on
$\omega_\ts$ and $\omega_\ti$, respectively. Motivated by the
$\delta$-function in \refeqn{Eqn:hamiltoniantime2}, we let
$\omega_\ts=\omega_{0\ts} + \epsilon$ and $\omega_\ti =
\omega_{0\ti} - \epsilon$, and make a series expansion of the
k-vectors:
\begin{subequations}
  \label{Eqn:kSeries}
  \begin{align}
    \label{kSeries}
    k_\ts & \approx k_{0\ts} + \epsilon\frac{\ud k_{0\ts}}{\ud \omega_{0\ts}} =
    k_{0\ts} + \epsilon\frac{1}{v_{\tg,\ts}^Z} = k_{0\ts}
    + \epsilon\frac{n_{\tg,\ts}^Z}{c}\\
    k_\ti & \approx k_{0\ti} - \epsilon\frac{\ud k_{0\ti}}{\ud \omega_{0\ti}} =
    k_{0\ti} - \epsilon\frac{1}{v_{\tg,\ti}^Z} = k_{0\ti} -
    \epsilon\frac{n_{\tg,\ti}^Z }{c}.
  \end{align}
\end{subequations}
In a spherical coordinate system, we have $p =
\sin\theta\cos\varphi$, $q = \sin\theta\sin\varphi$, and $m =
\cos\theta$, and so the phase-mismatch vector components become
\begin{align}
  \Delta k_x &= k_\ts \sin\theta_\ts\cos\varphi_\ts + k_\ti
  \sin\theta_\ti\cos\varphi_\ti \approx 0,
  \nonumber\\
  \Delta k_y &= k_\ts \sin\theta_\ts\sin\varphi_\ts + k_\ti
  \sin\theta_\ti\sin\varphi_\ti \approx 0,
  \nonumber\\
  \Delta k_z &= k_\ts \cos\theta_\ts + k_\ti \cos\theta_\ti - k_{0\tp}
  + K \nonumber \\
  & \approx \frac{\epsilon}{c}(n_{\tg,\ts}^Z - n_{\tg,\ti}^Z),
  \label{Eqn:mismatchVectorComponents}
\end{align}
where we have done a first-order approximation of $\sin\theta$ and
$\cos\theta$ for small angles, meaning that we consider only plane
waves, and where the last component is simplified using the
phase-matching condition for the forward direction, $k_{0\ts} +
k_{0\ti} - k_{0\tp} + K = 0$, together with \refeqn{Eqn:kSeries}.
Thanks to \refeqn{Eqn:mismatchVectorComponents}, we can now
trivially perform the integration over the spatial modes $\ud
\theta_\ts, \ud \theta_\ti$ and $\ud \varphi$, which finally leads
to the following compact expression
\begin{align}
  \frac{1}{i\hbar} \int\!\! \ud t \hat H(t)
  & = \frac{1}{i\hbar} \chi_2^\ph\: f_1 E_0 \nonumber\\*
  & \quad \times\: \int\!\!\ud \epsilon\: A_\ts(\epsilon) A_\ti(\epsilon)
  \hat{a}_\ts^\dag(\epsilon)
  \hat{a}_\ti^\dag(\epsilon) \nonumber\\*
  & \quad \times\: L\: \text{sinc}\left[\frac{L \epsilon}{2c}(n_{\tg,\ts}^Z -
    n_{\tg,\ti}^Z)\right] \nonumber\\*
  & \quad -\: \mathrm{H.c.} \nonumber\\*
  & = \int\!\!\ud \epsilon\: U(\epsilon)\: \hat{a}_\ts^\dag(\epsilon)
  \hat{a}_\ti^\dag(\epsilon)\: -\: \mathrm{H.c.}.
  \label{Eqn:hamiltoniantime3}
\end{align}

In summary, \refeqn{Eqn:evolution}, via
\refeqn{Eqn:hamiltoniantime3}, has helped us find the frequency and
polarization state generated in one crystal, which we will write in
the form
\begin{align}
  \ket{\Psi_{ZZ}} & = \frac{1}{B} \int\!\! \ud \epsilon\: U(\epsilon)\:
  \ket{\bm{\epsilon}} \otimes \ket{\bm{\chi}_{ZZ}^\ph},
  \label{Eqn:onecrystalfinalstate}
\end{align}
where $U(\epsilon)$ is defined by \refeqn{Eqn:hamiltoniantime3}, and
where
\begin{align}
  B & = \frac{\left(\int\! \ud \epsilon |A_\ts(\epsilon)
      A_\ti(\epsilon)|^2\: \text{sinc}^2[L \epsilon (n_{\tg,\ts}^Z -
      n_{\tg,\ti}^Z)/2c]\right)^{1/2}}{\hbar (\chi_2^\ph\: f_1
    E_0 L)^{-1}_\ph},
\end{align}
is a normalization constant, such that $|\frac{1}{B} \int\! \ud
\epsilon\: U(\epsilon)|^2 = 1$. Here, $\bm{\epsilon}$ represents the
frequency mode and $\bm{\chi}_{ZZ}^\ph$
represents the polarization mode along the $Z$-axis.\\

\pagebreak


\end{document}